\documentclass[conference]{IEEEtran}
\IEEEoverridecommandlockouts
\usepackage{cite}
\usepackage{graphicx}
\usepackage{hyperref}
\begin{document}

\title{AVHYAS: A Free and Open Source QGIS Plugin for Advanced Hyperspectral Image Analysis\\
}

\author{\IEEEauthorblockN{Rosly Boy Lyngdoh\IEEEauthorrefmark{1},
		Anand S Sahadevan\IEEEauthorrefmark{2}, Touseef Ahmad,
		Pradyuman Singh Rathore,\\ Manoj Mishra, Praveen Kumar Gupta and  Arundhati Misra}
	\IEEEauthorblockA{Hyperspectral Techniques Development Division,\\ Advanced Microwave and Hyperspectral Techniques Development Group,\\ Space Applications Centre, ISRO, Ahmedabad, Gujarat, India \\ Email: \IEEEauthorrefmark{1} roslylyngdoh@sac.isro.gov.in, \IEEEauthorrefmark{2}anandss@sac.isro.gov.in}}
\maketitle

\maketitle

\begin{abstract}
Advanced Hyperspectral Data Analysis Software (AVHYAS) plugin is a Python-3 based Quantum-GIS (QGIS) plugin designed to process and analyse hyperspectral (Hx) images. Starting with version 1.0, AVHYAS serves as a free and open-source platform for sharing and distributing Hx data analysis methods among research scholars, scientists and potential end-users. It is developed to guarantee full usage of present and future Hx airborne or spaceborne sensors and provides access to advanced algorithms for Hx data processing. The software is freely available and offers a range of basic and advanced tools such as atmospheric correction (for airborne AVIRIS-NG image), standard processing tools as well as powerful machine learning and Deep Learning interfaces for Hx data analysis. This paper gives an overview of the AVHYAS plugin, explains typical workflows and use cases for making it a constantly used platform for hyperspectral remote sensing applications. 
\end{abstract}


\section{Introduction}
Hyperspectral Remote Sensing (HRS) is a powerful remote sensing approach for detecting and monitoring the biophysical characteristics of the Earth’s surface. Successful applications with HRS include monitoring agricultural areas, forests, urban areas, snow and ice, atmosphere, inland waters, oceans, and other natural ecosystems. Wide varieties of methods are used for analysing the hyperspectral data \cite{bioucas2013hyperspectral} (e.g., classification, target Physico-chemical property estimation, target abundance estimation, radiative transfer modelling etc.). Some Hx sensors onboard airborne and spaceborne platforms have provided information-rich datasets with high spectral and spatial resolution. However, the methods used in multi-spectral (Mx) remote sensing data analysis are not adaptable for hyperspectral data processing due to the high collinearity (redundant information) between the adjacent bands. Other difficulties are the high dimensionality of the hyperspectral data, spectral mixing, systematic or non-systematic noises and atmospheric effects. Therefore, data mining from Hx image needs complex and highly compute-intensive data processing and analysis algorithms for various applications. Hyperspectral imaging systems such as AVIRIS-NG, Hyperion, PRISMA, HySIS and future hyperspectral missions (GISAT-ISRO, EnMAP-DLR, HyspIRI-NASA) along with in situ measurements from spectroradiometer, will provide ample scope of scientific studies, analysis, and value-added products generation for scientific missions. 
\begin{figure}
	\centering
	\includegraphics[scale=.25]{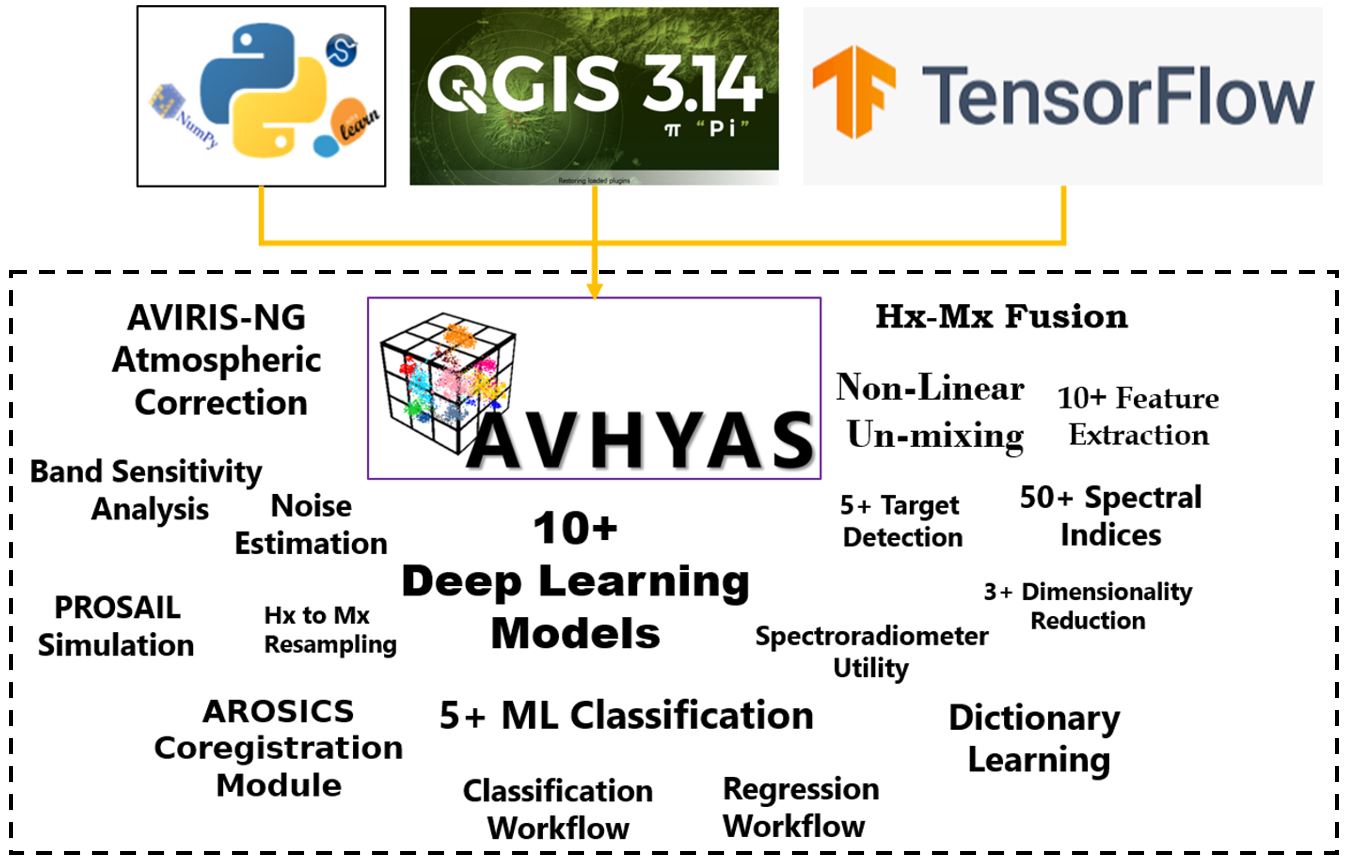}
	\caption{Schematic of different functionalities of advanced hyperspectral data analysis plugin (AVHYAS)}
	\label{fig:figure0}
\end{figure} 
\begin{figure*}[t]
	\centering
	\includegraphics[scale=.5]{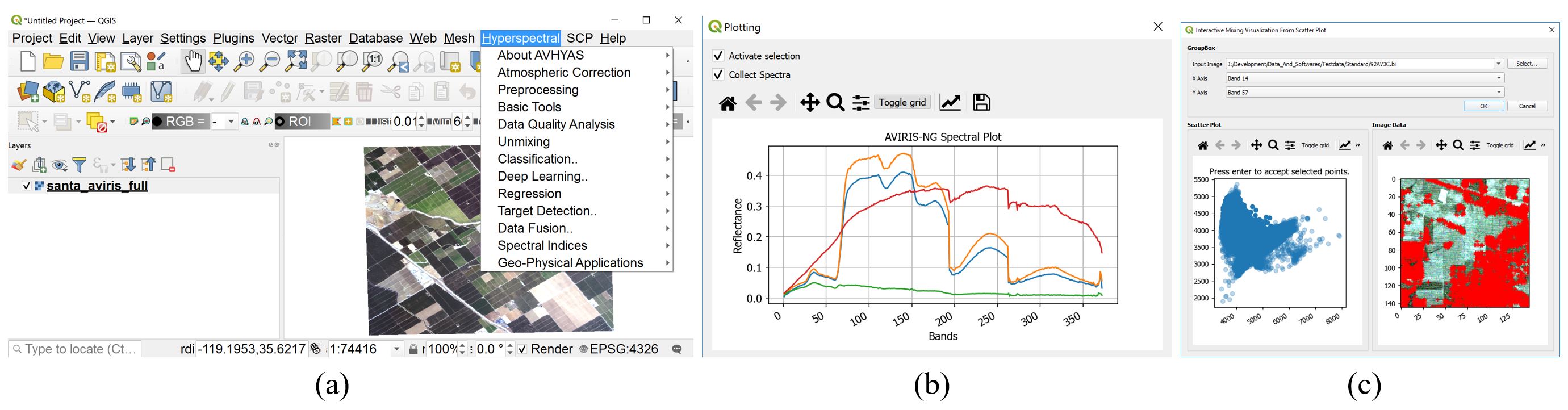}
	\caption{(a) AVHYAS plugin is integrated with QGIS as 'Hyperspectral' Menu, (b) spectral viewer, (c) scatter Plot viewer for finding the pure and mixed pixels}
	\label{fig:figure1}
\end{figure*} 

Hyperspectral data analysis requires specific algorithms to analyze such extensive dimensional data (e.g. parallelopiped classification, broad-band ratios and multi-linear regression, which are not suited for analyzing Hx data). Widely used Mx data processing software requires additional functionalities (e.g. feature selection and feature extraction) to exploit the wealth of rich spectral information. At present, advanced algorithms such as deep learning (DL), sparse-linear un-mixing are not available in commercial or open-source software. Moreover, advanced algorithms which are available in different scripting languages (usually written in C/C++, Java, Python and Matlab) do not have a simple user interface (UI) for the users having no prior programming skill \cite{abadi2016tensorflow}. Nevertheless, libraries for advanced algorithms are usually domain-specific and are not provided with user interfaces. Therefore, programming skills beyond that of Mx data users are required to perform computationally intensive tasks for advanced algorithms \cite{audebert2019deep},\cite{ahmad2020four}. Over the last decade, Python scripting language opened up new and simple pathways for effective implementation of machine learning (ML) and DL algorithms, making it possible to perform the programming tasks in a few lines of code that would otherwise take hundreds of lines in lower-level languages. The success of Python scripting languages is vastly based on the constant improvement of shared scripts and resources within an active community. In this context, the concept for the AVHYAS was developed with a focus on exploring the full potential of QGIS and other powerful python modules for Hx data analysis (see Fig. \ref{fig:figure0}). The AVHYAS plug-in efficiently bridges all advantages of QGIS (e.g., visualization, vector data processing) and powerful python packages like GDAL (for data I/O or working with virtual raster files), TensorFlow, scikit-learn, scikit-image for Hx data processing and analysis in Windows-based operating system. It also uses QGIS-interface and other python packages such as matplotlib, seaborn to visualize Hx data. 

The Overarching aim of AVHYAS is to provide free-of-charge and user-friendly access to advanced approaches for Hx data processing for both beginners and advanced users. The AVHYAS is integrated into the QGIS classic menu to extend its range of available applications. It can also be used with any Mx imagery for specific applications (e.g., classification, regression, feature extraction etc.). A standard workflow was adopted for effectively integrating machine learning approaches to the QGIS environment. This way, functionalities were implemented that include the standard methods available in other proprietary software (e.g., ENVI, ArcGIS) or non-commercial/open-source software (e.g., EnMAP-Box \cite{rabe2018enmap}) and the advanced algorithms (e.g. DL based classification) which are not available in the standard software. This paper provides an overview of the AVHYAS and examples for some of the use cases (technical documentation of the AVHYAS plugin is included in the installation bundle available at the website (\url{https://sites.google.com/view/avhyas-sac-isro/home}).

\section{AVHYAS Framework}
The AVHYAS can be added to the QGIS as a plugin (see Fig. \ref{fig:figure1} (a)) and shall be registered in the QGIS plugin repository. General tasks like raster and vector file management follow the same principles and look-and-feel available in QGIS. These functionalities are extended based on more specific requirements for Hx data (e.g. spectral viewer, RGB-tool for visualizing true colour composite (TCC) and false colour composite (FCC)) or the management of Hx data analysis workflow (e.g. text view for presenting classification accuracy in HTML format). Views can interact in various ways with each other; for example, the selected image spectral profile can be visualized in a spectral view together with multiple target spectra of the same image or different images. The AVHYAS uses generic file formats (available in GDAL) for storing image data (e.g., binary spectral data in band-sequential-order with metadata information). The header file is compatible with the widely used ENVI file format and the generated output, therefore compatible with other software products. An additional 'Add Data' button is included in the Hxtools RGB toolbar, which enables the fast loading of large Hx images without the rotational information. AVHYAS provides typical interactive tools for feature space visualization and spectral visualization. Multiple pixel spectra can be added to the spectral viewer to compare the spectral features (see Fig. \ref{fig:figure1} (b)). Interactive scatter plot visualization is provided with the un-mixing module to locate the pixels that represent the end member spectra or mixed spectra (see Fig. \ref{fig:figure1} (c)). A check box is provided with each module to load the generated image into QGIS canvas. User can use QGIS raster-functionalities to compute image band statistics and histograms. 

Besides data source management and visualization modules, the AVHYAS consists of Atmospheric Correction Module, Basic Tools Module, Pre-processing Module, Data Quality Analysis Module, Un-mixing Module, Classification Module, Deep Learning Module, Regression Module, Fusion Module, Spectral Indices, and Geo-Physical Applications Module. 

\section{AVHYAS Tools and Applications}
\subsection{Atmospheric Correction Module for AVIRIS-NG Data}
Applications of high-spatial-resolution imaging spectrometer data acquired from the
AVIRIS-NG require a thorough compensation for atmospheric absorption and scattering. Atmospheric Correction utility is for converting Level-1 radiance image of advanced visible-near-infrared-imaging-spectroradiometer next-generation (AVIRIS-NG) airborne-sensor to Level-2 Reflectance image. Retrieval of water vapour and aerosol optical depth (AOD) over land is critical for correcting atmospheric influence on Hx images. The dark dense vegetation method and radiative transfer modelling are used to derive AOD. Estimation of perceptible water vapour is carried out using short-wave hyperspectral measurements for each pixel. A differential absorption technique (continuum interpolated band ratio) has been used for this purpose \cite{mishra2019retrieval}.
Further, these parameters were used to derive ‘atmospherically corrected surface reflectance for land pixels, assuming horizontal surfaces having Lambertian reflectance. This module contains an entirely standalone version of the operational model to convert
radiance data to reflectance data \cite{mishra2019retrieval}. All the necessary auxiliary looks up tables have been bundled along with AVHYAS. 
\begin{figure}
	\centering
	\includegraphics[scale=.33]{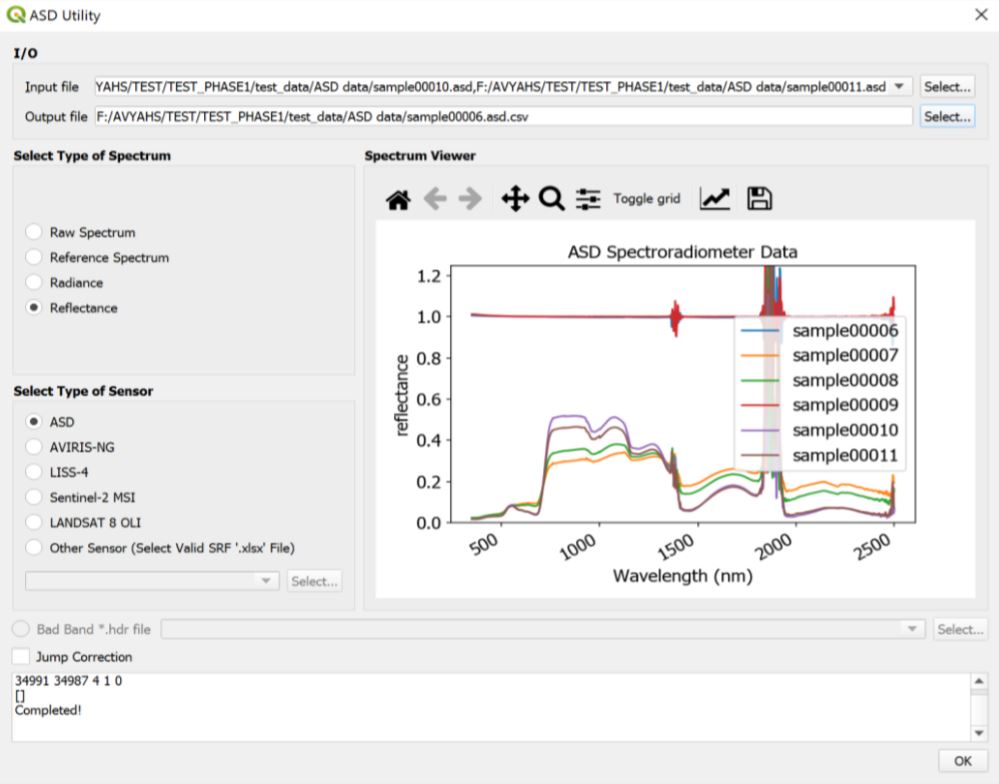}
	\caption{UI of ASD utility Module for visualising and resampling ASD spectroradiometer data}
	\label{fig:figure2}
\end{figure} 
\begin{figure}
	\centering
	\includegraphics[scale=.54]{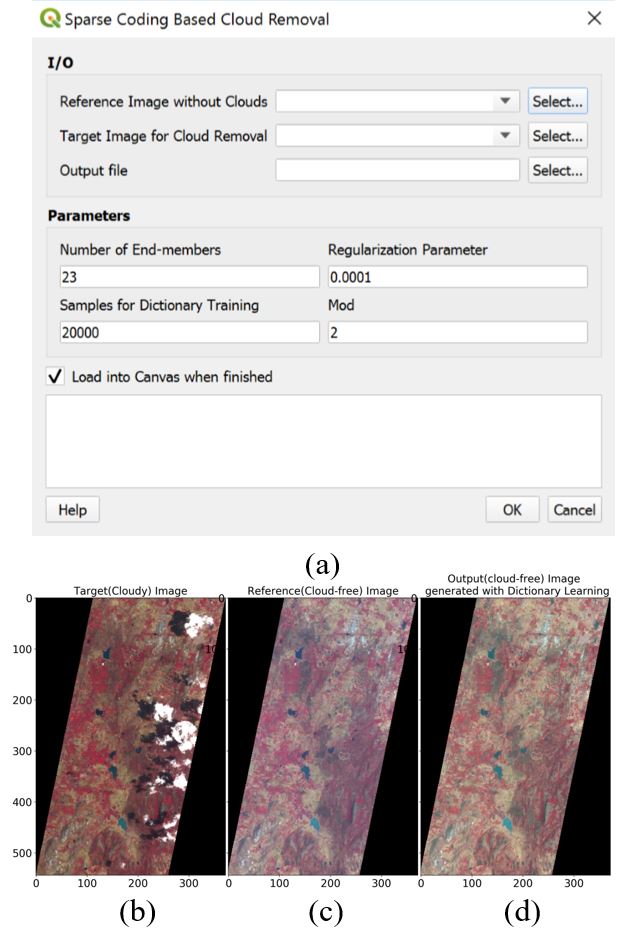}
	\caption{(a) UI of Cloud removal module, (b) target image with cloud (c) reference cloud-free (d) cloud removed target image}
	\label{fig:figure3}
\end{figure}
\begin{figure}
	\centering
	\includegraphics[scale=.38]{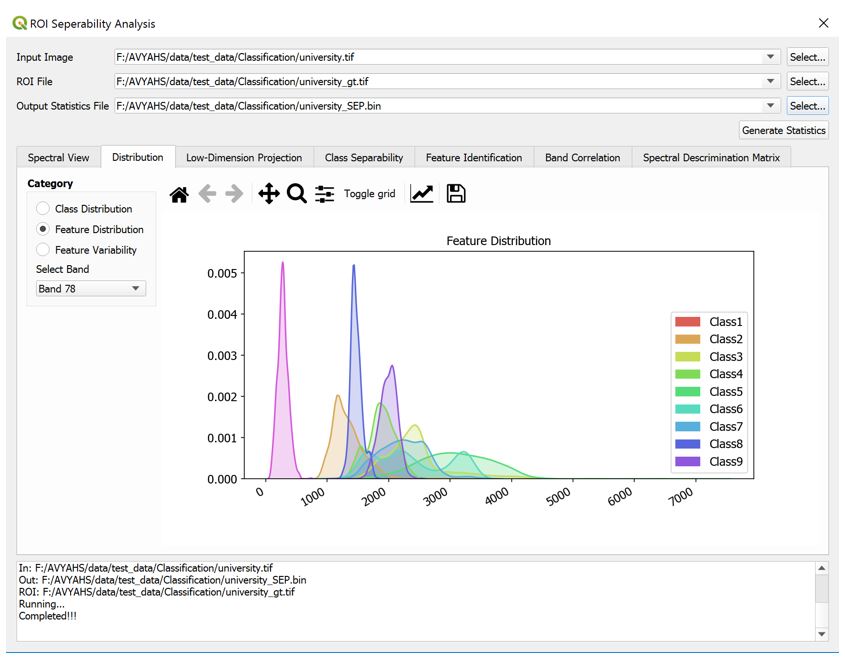}
	\caption{UI of ROI-separability analysis tool}
	\label{fig:figure4}
\end{figure} 

\subsection{Basic Tools Module}
Basic-Tools module contains sub-modules such as sensor-utility, data-subset, spectral plot, scatter plot, scaling, and ROI Separability. Sensor utility module contains four sub-modules to read ASD spectroradiometer data (Fig. \ref{fig:figure2} illustrates the UI of ASD utility Module for visualizing and resampling ASD spectroradiometer data, to generate spectral-response-function (SRF) from full-width-half-maximum (FWHM) and the central wavelengths of the sensor, to perform spatial and spectral resampling of the Hx images to generate Mx images. Scaling functionality allows the user to scale the Hx images by a factor to bring the Hx data within a specific numerical range. ROI separability module is designed to perform the class separability analysis and feature separability analysis to identify the separable classes, visualize the mean and standard deviation spectra of different classes, low-dimensional visualization of different classes, and identify the optimum bands to discriminate different classes.  
\begin{figure}
	\centering
	\includegraphics[scale=.33]{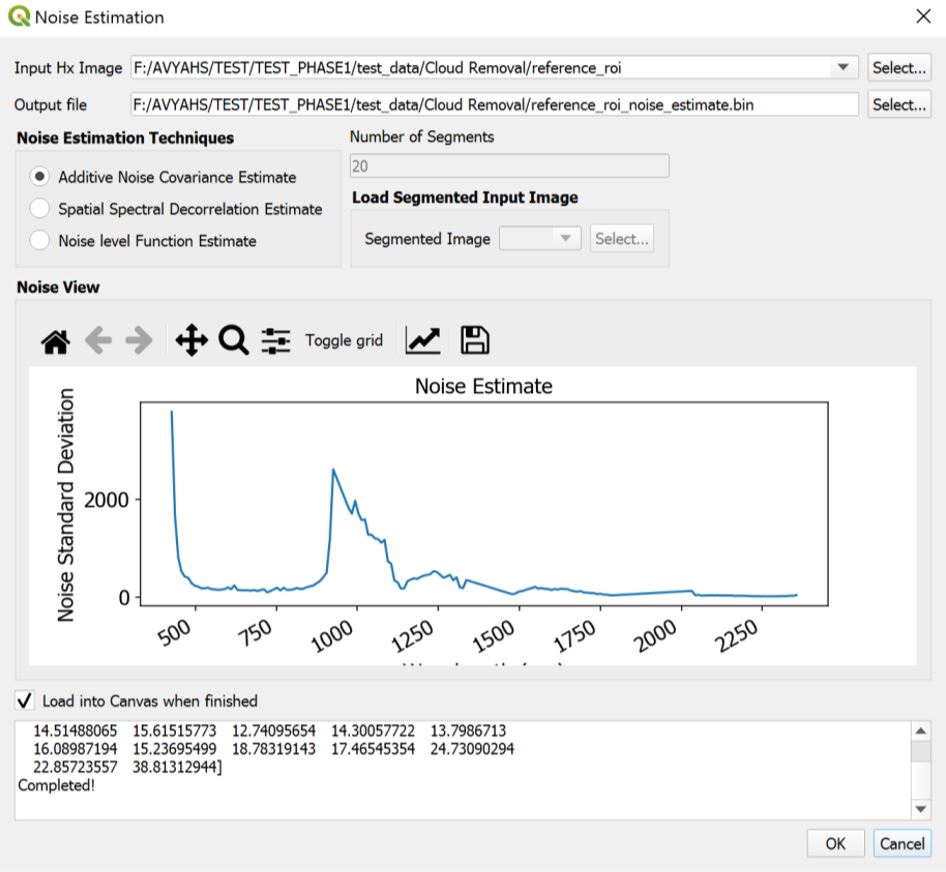}
	\caption{UI of noise estimation module to visualise the band-wise noise standard deviation}
	\label{fig:figure5}
\end{figure}
\begin{figure}
	\centering
	\includegraphics[scale=.26]{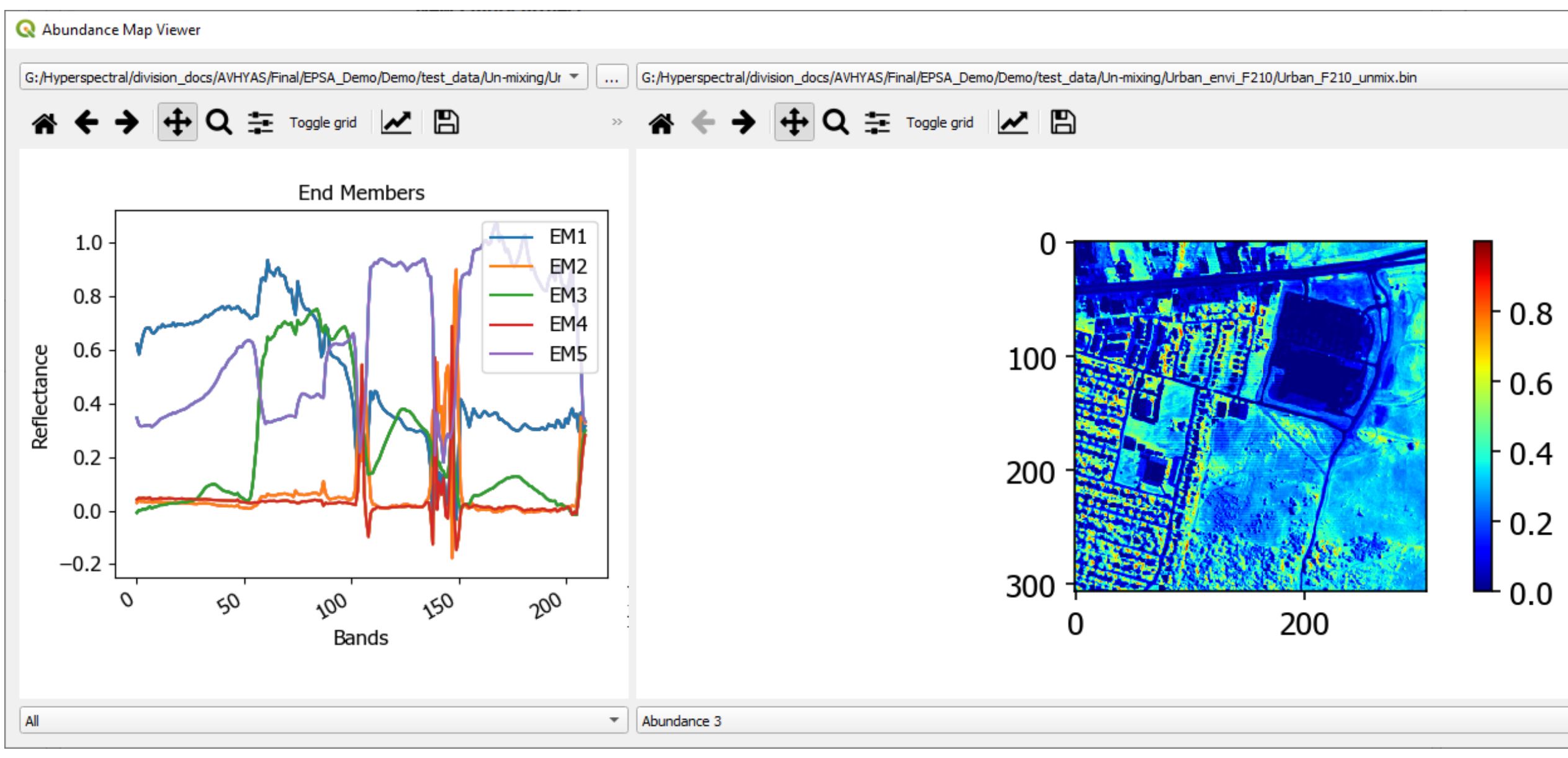}
	\caption{UI of abundance map viewer to visualise the un-mixing results}
	\label{fig:figure6}
\end{figure}
\begin{figure*}[t]
	\centering
	\includegraphics[scale=.48]{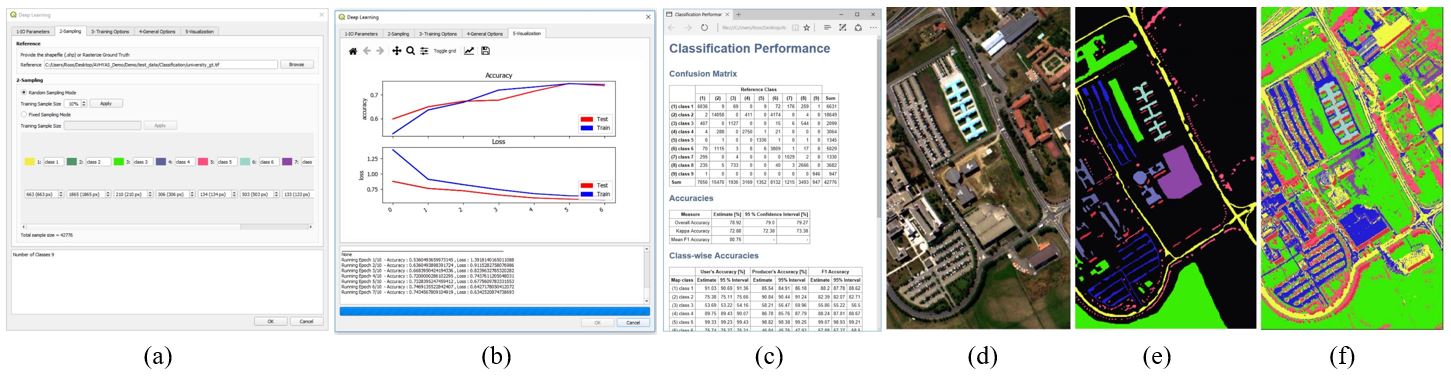}
	\caption{(a) UI of deep learning based classification workflow (b) inference visualisation for Deep Learning (c) report of classification accuracy displayed on web-browser using HTML (d) Pavia University Hx image, (e) ground truth of the image, (f) classified map generated using AVHYAS 3D-CNN model}
	\label{fig:figure7}
\end{figure*} 
\subsection{Pre-processing and Data Quality Analysis Modules}
Pre-processing module contains sub-modules such as Dimensionality Reduction (DR), General Purpose Utility, Feature Extraction, Data Transformation and Savitsky-Golay-Filtering. Dimensionality reduction (DR) \cite{khodr2011dimensionality} utility can perform linear (e.g. minimum-noise-fraction, MNF) and non-linear DR (e.g. Kernel PCA) on Hx images.
Cloud removal \cite{xu2016cloud} tool in General-Purpose-Utility module can perform cloud removal in Hx images using sparse-coding and dictionary learning. Figure \ref{fig:figure3} (a) illustrates the UI of the cloud removal module, Fig.\ref{fig:figure3}(b) shows the cloudy image (EO1-Hyperion image), (c) shows the reference cloud-free image obtained from the Hyperion sensor, and (d) illustrates the result obtained using the cloud removal tool (cloud removed target image).  
Feature extraction employs local spectral similarity \cite{sahadevan2019spatial}, spatial-spectral-gradient and extended morphological filter \cite{plaza2005dimensionality} algorithm to extract spatial features from the Hx image. Similarly, the fit-transform module can perform different data transformations such as standard-scalar \cite{pedregosa2011scikit}, min-max scalar, robust-scalar transformations, and continuum-removal \cite{gomez2008continuum} to enhance the spectral features in the Hx images, which are useful pre-processing functionalities to improve the performance of classification and regression algorithms. 

Data Quality Analysis module has been developed to analyse the quality of the Hx images in terms of spectral and spatial characteristics. This module has seven sub-modules such as signal-to-noise ratio (SNR), bad-band detection, whitening, noise estimation \cite{gao2013comparative}, and de-striping. Noise estimation module can be applied to analyse the spectral-noise, spatial-spectral-noise and to estimate the noise in the spectrally homogeneous regions (Fig.\ref{fig:figure6} illustrates UI of the noise estimation tool and the plot showing the noise-standard-deviation obtained from the radiance image). The bad-band detection method requires a threshold value to demarcate the noisy channel based on the user-defined threshold.  
\subsection{Un-mixing Module}
The field of spectral mixture analysis (or spectral unmixing) is dedicated to both identifying the most probable set of pure pixels (called endmembers) and estimating their proportions (called abundances) in each of the image pixels. The Un-mixing module performs the endmember extraction and the abundance estimation in Hx images. Sub-modules consist of Material-Count, endmember extraction, abundance estimation, sparse based un-mixing, interactive scatter plot visualization, visualization of un-mixing results, and un-mixing error analysis. The endmember extraction module includes the widely used algorithms such as PPI, NFINDER, ATGP and VCA \cite{chang2016comparative}. The abundance estimation module includes linear algorithms and generalized bi-linear models \cite{bioucas2012hyperspectral}. The interactive scatter plot visualization module provides an interface for visualizing the un-mixing results. QGIS, as such, has no viewer for visualizing spectra and abundances. This utility has been implemented so as the user can visualize the different endmembers along with their respective abundances and generate the root-mean-square-error (RMSE) map of the abundance estimation methods (Fig. \ref{fig:figure6} illustrates the UI of abundance map viewer to visualize the un-mixing results).

\subsection{Classification and Deep Learning Modules}
The classification and Deep Learning module are designed to provide simple user interfaces to perform classification of Hx images. Classification module consists of supervised classification, unsupervised classification and segmentation modules. Supervised classification module can perform classification of Hx images using widely used algorithms such as spectral angle mapper, support-vector-machine, random-forest etc. These modules employ scikit-learn package \cite{pedregosa2011scikit} to perform the classification tasks. Moreover, the workflow functionality of these modules can perform hyperparameter tuning of the specific algorithm to achieve better accuracy. DL module consists of state-of-the-art DL algorithms for the classification of Hx images \cite{audebert2019deep}, \cite{paoletti2019deep}. The inference sub-module of DL provides an interface for performing prediction based on a trained deep learning model. The user has to provide a valid H5 file \cite{abadi2016tensorflow} which contains the weight and biases of the model being trained (using AVHYAS). The design of the UI for classification workflow has been adopted from the EnMap-Box toolbox. Model performance is printed as html reports and the assessment report will be pop up in the default browser. The segmentation module consists of spatial-spectral, split-merge and mean-shift segmentation methods, which can be used for segmenting the spatial-spectral homogeneous patches in Hx images. 

Pavia University scene was used for generating the output of the DL module. This data was acquired by the Reflective Optics System Imaging Spectrometer(ROSIS-3) over the city of Pavia, Italy (see \cite{yokoya2011coupled}, and \cite{paoletti2019deep} for the detailed information about the dataset). Figure \ref{fig:figure7} (a), (b) and (c) illustrate the UI of deep Learning-based Classification Workflow, inference visualisation and classification performance report, respectively. Figure \ref{fig:figure7} (d), (e) and (f) show the Pavia University Hx image (TCC), the ground truth of the image and the classified map generated using the 3D-CNN model available in AVHYAS.

\begin{figure}
	\centering
	\includegraphics[scale=.45]{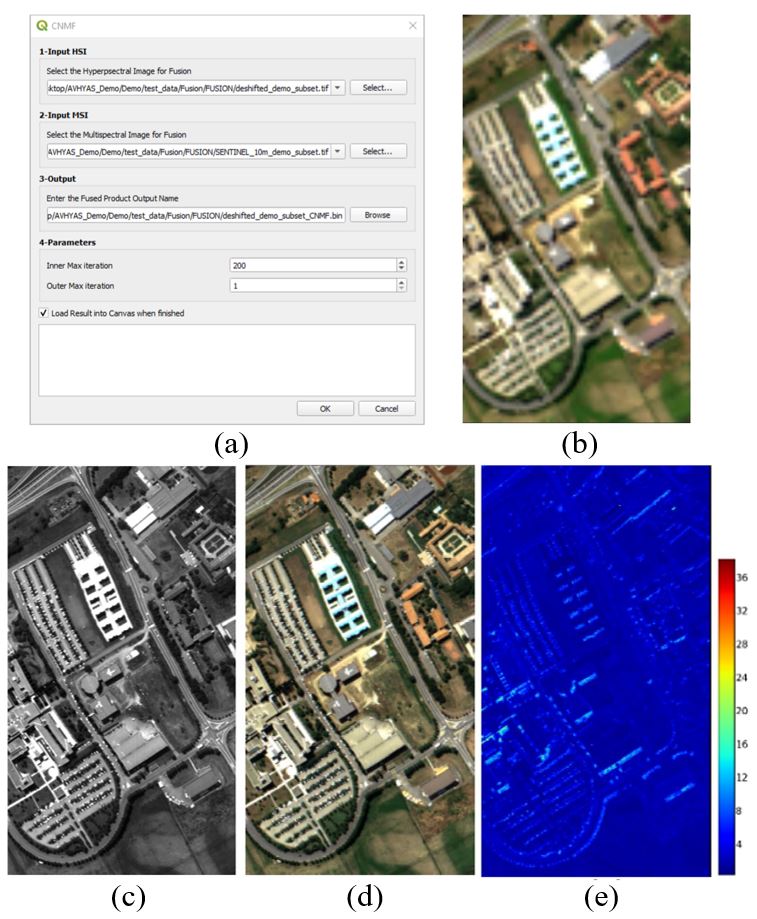}
	\caption{(a) UI of Hx-Mx image fusion module, (b) synthetic low spatial resolution Pavia University Hx image, (c) synthesize high spatial resolution Mx data over Pavia Univ. (d) High spatial resolution Hx image generated using the fusion module (e) Spectral-Angle-Mapper error between the reconstructed and the original image}
	\label{fig:figure8}
\end{figure}
\subsection{Regression, Fusion, Geo-Physical Applications and Spectral Indices Modules}
Regression is a statistical method used in remote sensing to build a relationship between the desired attributes (Physico-chemical or biological properties of target) and their Hx signature. In AVHYAS, the regression module provides a flexible way of running different regression algorithms (e.g. PLSR, Support-vector-regression, random-forest-regression, etc.) on Hx images \cite{torres2020overview}. The idea of workflow has been borrowed and inspired from Enmapbox plugin \cite{van2015enmap} with the back-end module written and prepared in house. 

At present, Geo-Physical Applications Module consists of an interactive simulation tool for generating the vegetation spectra using the PROSAIL model. The idea of workflow has been borrowed from Enmapbox plugin \cite{van2015enmap} with the additional sensor resampling functions for Mx sensors of Indian-Space-Research-Organisation (ISRO) was incorporated. Similarly, the vegetation spectral indices module has been adopted from the Enmap toolbox \cite{van2015enmap}. Moreover, the future versions of AVHYAS will have the spectral index module for other targets such as soil, water etc. 

Data Fusion module in AVHYAS is designed for enhancing the hyperspectral spatial resolution using high spatial-resolution Mx image. AROSICS (Automated and Robust Open-Source Image Co-Registration Software), a python-based open-source module \cite{scheffler2017arosics} was integrated with AVHYAS for automatic detection and correction of sub-pixel misalignment between Hx and Mx images. This module can improve the co-registration between Hx and Mx images and may substantially improve the data fusion performance \cite{yokoya2011coupled}. Figure \ref{fig:figure8} (a) shows the UI of data fusion module. Figure \ref{fig:figure8} (b) and (c) show the synthesized low spatial resolution (10m) Pavia University Hx image and the high spatial resolution Mx image (5m), respectively. Figure \ref{fig:figure8} (d) and (e) show the high spatial resolution Hx image generated using the fusion module, and pixel-wise cosine distance \cite{yokoya2011coupled} between the fused image and the original image. 

AVHYAS is not limited to HRS data of the airborne or spaceborne sensors, and it can also be used for analyzing the data acquired by the handheld/tabletop Hx cameras. Moreover, the chance for the AVHYAS to become an evolving plugin with a constantly growing set of applications is high, given its flexibility of integrating new algorithms for different Hx sensors and new powerful ML/DL libraries.

\section{Conclusion}
AVHYAS integrates powerful machine-learning and deep-learning algorithms to perform various data analysis tasks for extracting information from Hx data. Basic and advanced algorithms were incorporated in the AVHYAS toolbox aiming at the extension of the user community (in the field of HRS) in India and providing the most powerful algorithms for the analysis of the present and the future Hx-imaging sensor data. The AVHYAS plugin development was thus driven by the idea of familiarising advanced Hx image analysis algorithms with the multi-disciplinary community. It was used for training the academia and research community to get hands-on experience on Hx image analysis and always received positive responses. A set of multi-disciplinary applications will be integrated into future versions, such as non-linear un-mixing for mineral mapping, soil spectral indices, Look-up table-based inversion for biophysical parameter estimation, Deep-Learning based regression, Ensemble classification for multi-modal data etc. Moreover, the future versions of AVHYAS will have parallel implementations of a widely used hyperspectral data analysis algorithm, which will help users process large-sized Hx images on multi-core processors.

\section*{Acknowledgment}
The AVHYAS plugin is developed at Hyperspectral techniques Development Division, Space Applications Centre (SAC), Indian Space Research Organisation (ISRO), Ahmedabad, Gujarat, as part of the Microwave and Hyperspectral Techniques for Earth Resources Applications and Management (MAHTRAM) project. The authors are grateful to the technical evaluation committee members in SAC for their valuable feedback, which helped us improve the functionalities of AVHYAS.
AVHYAS installation bundle available at the website (\url{https://sites.google.com/view/avhyas-sac-isro/home})

\bibliographystyle{IEEEtran}
\bibliography{bibliography}

\end{document}